\documentclass[preprint,showkeys,nofootinbib,preprintnumbers,amsmath,amssymb,floatfix]{revtex4-1}

\usepackage{graphicx}
\usepackage{color}
\usepackage{epsfig}

\newcommand{\bea}{\begin{eqnarray}}
\newcommand{\eea}{\end{eqnarray}}
\newcommand{\vect}[1]{\mathbf{#1}}

\newcommand{\kt}{k_{\rm B}T}
\newcommand{\nveq}{n_{\rm v,eq}}
\newcommand{\dgv}{\Delta G_{\rm v}}
\newcommand{\nmax}{n_{\rm m}}

\newcommand{\eq}[1]{eq.~(\ref{#1})}

\newlength\picwidth
\begin{document}

\title{Thermal vacancies in close--packing solids}

\author{M.~Mortazavifar and M. Oettel} 
\affiliation{Institut f\"ur Angewandte Physik, Universit\"at T\"ubingen,
  Auf der Morgenstelle 10, 72076 T\"ubingen}

\date{\today}

\begin{abstract}
	Based on Stillinger's version of cell cluster theory, we derive an expression for the equilibrium concentration
	of thermal monovacancies in solids which allows for a transparent interpretation of the vacancy volume and the
	energetic/entropic part in the corresponding Gibbs energy of vacancy formation $\dgv$. For the 
	close--packing crystals of the hard sphere and Lennard--Jones model systems very good agreement with simulation
	data is found. Application to metals through the embedded--atom method (EAM) reveals a strong sensitivity   
	of the variation of $\dgv$ with temperature to details of the EAM potential. 
	Our truncation of the cell cluster series allows for an approximate, but direct measurement of crystal free energies and 
	vacancy concentration in colloidal model systems using laser tweezers.
\end{abstract}


\maketitle

{\em Introduction.--}
	Point defects, in particular vacancies, are the simplest deviation from an ideal 
	crystal with perfect translational symmetry. At finite temperature, the thermal motion of
	the atoms in a crystal necessarily create vacancies 
	and causes a finite equilibrium concentration $\nveq$ of these. Thus $\nveq$ 
	is a very basic property of a crystalline material (with significance for
	other properties, most notably self--diffusion), nevertheless the magnitude and temperature dependence
	of $\nveq$ have been debated for nearly a century \cite{Kra00} without a clear consensus reached.  
	It is customary to express the equilibrium vacancy concentration as
	$\nveq=\exp(-\beta \dgv)$ with $\dgv$ being the    
	Gibbs energy of vacancy formation and $\beta=1/(\kt)$ is the inverse temperature.
	In the materials science community, $\dgv$ is usually further analyzed in terms of its
	entropic and enthalpic part where the entropic part is the more difficult one: here, the effects
	of lattice vibrations distorted by vacancies and anharmonicity enter. For metals, even approximate calculations of the
	entropic part using quantum density functional theory are a case for supercomputers and beset with 
	uncertainties \cite{Neu11,Neu12}. On the other hand, coming from classical statistical physics one would
	assume that the particles forming the crystal interact with
	classical potentials. Then the problem of vacancies and other point defects 
	can be tackled using methods from
	statistical mechanics. Quantitative results have been obtained mainly by Monte Carlo simulations, and
	here with a focus on hard spheres (HS) \cite{Ben71,Pro01,Kwa08}. A deeper understanding of these results 
	is nevertheless missing. The equally important Lennard--Jones (LJ) model system 
	has received less attention from the simulation side \cite{Jac80}.

	In this paper, we approach the problem from the theoretical side using an expansion of the crystal
	partition function developed by Stillinger and coworkers \cite{Sti65} (which falls into the realm of
	cell cluster theories).
	The expansion parameter is the number $n$ of contiguous particles (free to move) in an otherwise frozen matrix of particles at their
	ideal lattice positions. For the lowest--order term in this series ($n=1$) we find a simple formula for $\dgv$
	in terms of the equation of state and some three--dimensional configuration integrals which are easy to calculate
	(see \eq{eq:master_formula} below).
	The agreement with simulation data in case of the two model systems HS and LJ is very good 
	and provides us with an interpretation on the different origin of a finite vacancy concentration in solids
	with purely repulsive interactions between atoms (such as HS) and in solids with attractive interactions (such as LJ).
	We proceed by examining Ni as an example for a metal with a likewise face-centered cubic (fcc) crystal structure
	using the method of embedded--atom potentials (EAM). It turns out that EAM potentials in the physical spirit 
	close to the original work \cite{Daw84} give a temperature behavior of the vacancy concentration close to the LJ solid and in line
	with the rare simulation results  whereas
	potentials that are used as mere fitting devices show a very different behavior.     

{\em Theory.--}
	Consider the canonical partition function $Q=Q(N,V,T)$ for $N$ atoms interacting with a classical potential
	$\phi(1...N)$ of many--body nature, depending on the individual atom positions $\vect r_1....\vect r_N$:
		\bea
		  Q &=& \frac{1}{N!\Lambda^{3N}} \int d\vect r_1 \dots \int d\vect r_N \exp(-\beta \phi(1...N) ) \;. 
		\eea
	Here, $\Lambda$ is the de Broglie wavelength. We consider a face--centered cubic (fcc) reference lattice with
	$M \ge N$ lattice sites at positions $\vect s_i$ spanning the volume $V$. We associate each particle $i$ with a lattice site
	at site $\vect s_i$ and that association divides the $3N$ dimensional configuration space into non overlapping
	regions $\Omega_{l,p}$. The precise form of this association is discussed in ref.~\cite{Sti65}, but one may think
	of it loosely in terms of each particle $i$ belonging to the Voronoi cell around site $\vect s_i$ of the lattice.
	For a chosen subset of $N$ lattice sites $\{\vect s_i\}$ and associated cells, the index $p$ runs over the $N!$ permutations
	of the particles among these cells and this  leads to an identical division of the configuration space,
	$\Omega_{l,p_1} \equiv \Omega_{l,p_2}$. The index $l$ runs over the different associations of $N$ particles with
	$M>N$ lattice sites.
	Thus we obtain for the partition function:
		\bea
		  Q &=& \frac{1}{\Lambda^{3N}} \sum_l \int \dots \int_{\Omega_{l,1}} \!\!\!d\vect r_1 \dots  d\vect r_N \exp(-\beta \phi(1...N) ) \;. \quad
		\eea
	Following ref.~\cite{Sti65}, one may write $Q$ in terms of configuration integrals $Z_i^l$, $Z_{ij}^l$, \dots which
	describe the correlated motion of one, two, \dots particles in a background matrix of $N-1$, $N-2$, \dots particles fixed at their
	associated lattice sites. These configuration integrals are defined as
		\bea
			Z_i^l		&=& \int_{\omega_i^l}		d\vect r_i \exp(-\beta \phi(1...N))		\;{\rm with} \;\vect r_j = \vect s_j \; (j\not =i) \;,	\nonumber\\
			Z_{ij}^l	&=& \int_{\omega_{ij}^l}	d\vect r_i d\vect r_j \exp(-\beta \phi(1...N))	\;{\rm with} \;\vect r_k = \vect s_k \; (k\not =i,j) \;,	\nonumber\\
					& \vdots & \;.
		\eea
	The integration domains must fulfill $\omega_{i}^l, \omega_{ij}^l, \dots \in \Omega_{l,1}$, and they depend on the indices of
	the free particles $i,j$ and also in the index $l$ determining at which lattice sites the other particles are fixed.
	The partition function is now expressed as the product
	\bea
	  Q &=& \frac{1}{\Lambda^{3N}} \sum_l \prod_i^N Z_i^l \; \; \prod_{i<j}^N \frac{Z_{ij}^l}{Z_i^l Z_j^l} \dots \;.
	   \label{eq:Qseries}
	\eea
	In the full product formula, the configuration integrals $Z^l_{i..}$ all cancel save for the $N$--particle configuration integral
	$Z^l_{1...N}$ which makes the last equation an identity \cite{Sti65}. The individual factors contain the effect of the (correlated)
	motion of one, two, ... particles in the crystal. 

\begin{figure}
			\centering
			\includegraphics[width=0.5\linewidth]{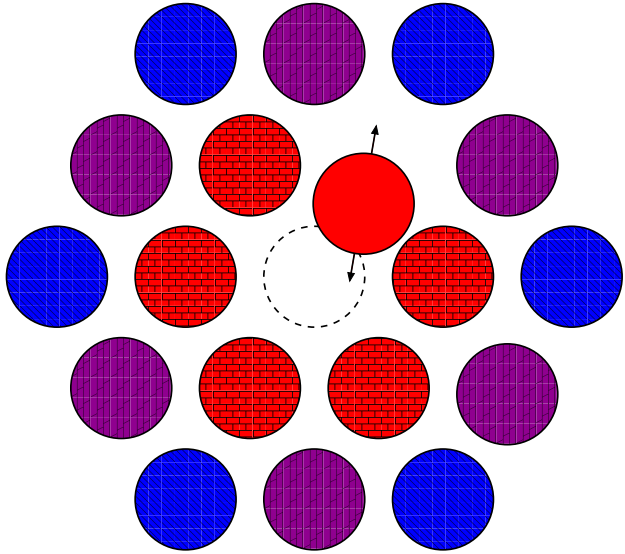}
			\caption{(color online) One free particle (circle with arrows) in the first shell neighbourhood of a vacancy with otherwise
			frozen particles. Different colors (shadings) mark different shells around the vacant position.}\label{fig:vacant_effect}
		\end{figure}

	We assume that the vacancy concentration is small, $n_{\rm v} = 1 - N/M \ll 1$, such that vacancies do not interact with each other.
	For the moment, 
	we truncate the product (\ref{eq:Qseries}) after the first term.
	The sum over $l$ (distribution of $N$ particles on $M$ sites) results in a factor $\binom{M}{N}$ and we find:
		\begin{equation}
			Q \approx  \frac{1}{\Lambda^{3N}} \binom{M}{N}
				\left( Z_s	\right) ^{N - (M-N)\sum_i^{\nmax} g_i}
				\prod_i^{\nmax}\left( Z_{sv,i}	\right) ^{g_i (M-N)} \;.
		\end{equation}
	Here, $g_i$ denotes the number of atoms in shell $i$ around a fixed lattice site (e.g. $g_1=12$, $g_2=6$,... in the isotropic HS/LJ systems
	with fcc structure). We assume a finite range of the interaction potential (up to shell $\nmax$) and therefore we have 
	for the $M-N$ vacancies a number of $(M-N)g_i$ of atoms which feel the vacancy at the shell $i$ apart. The single--particle 
	configuration integral for these particles is given by  $Z_{sv,i}$. All remaining particles are not influenced by the vacancies and
	contribute the single--particle integral $Z_s$. See also fig.~\ref{fig:vacant_effect} for a visualization.
	The equilibrium vacancy concentration is found by minimizing the free energy $\beta F = -\log Q$ with respect to $n_{\rm v}$. 
	A short calculation leads to 
		\bea
			\nveq \approx \exp\left( \frac{Z_s'}{Z_s}\rho_M   \right) \prod_i^{\nmax}\left( \frac{Z_{sv,i}}{Z_s}  \right)^{g_i} \;.  
		\eea
	Here, $\rho_M$ is the density of lattice sites ($\rho$ is the density of atoms) and $Z_s'=\partial Z_s/\partial \rho_M$. Since $\nveq \ll 1$, we find
	$\rho_M \approx \rho$ and $\beta F \approx -N \log Z_s$. Using the thermodynamic relation for the pressure $ p = (F/N)' \rho^2$ we find the
	following central result of our work for the Gibbs energy for vacancy formation:
		\bea
			\dgv = \frac{p}{\rho} - \beta^{-1}  \sum_i^{\nmax} g_i \log \left( \frac{Z_{sv,i}}{Z_s}  \right) \:.
			\label{eq:master_formula}
		\eea 
	The first term is just given by the crystal equation of state and ensures that $\nveq$ goes rapidly to zero for crystals under high pressure.
	Using the thermodynamic decomposition $\dgv = pV_{\rm v} + E_{\rm v} - T\Delta S_{\rm v}$ one sees that
	the ``vacancy volume'' $V_{\rm v}$ is just given by $1/\rho$. $E_{\rm v}$ and $- T\Delta S_{\rm v}$ are the energetic and
	entropic part and are contained in the second term in \eq{eq:master_formula} (which we call the vacancy integral term in the following).
	Although the splitting of $\dgv$ into equation--of--state and vacancy integral term was derived here in the leading truncation
	of the Stillinger series, the results stays valid in higher truncations (with further corrections to the vacancy integral term)
	{\em as long as the vacancies can be assumed to be noninteracting}.

\begin{figure}
				\includegraphics[width=\picwidth]{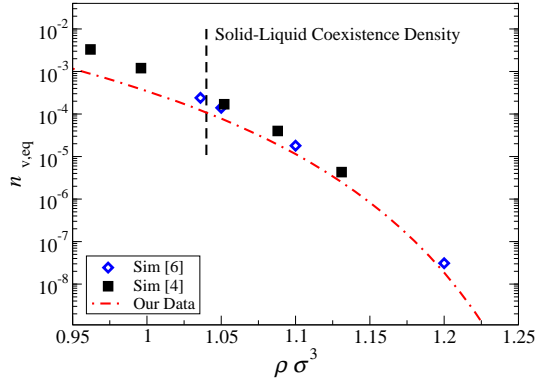}
				\caption{Hard spheres: vacancy concentration versus density. Error bars for the simulation data are unknown.}\label{fig:HS_nvac}
		\end{figure}

{\em Numerical results.--}
	For HS, the free energy per particle in the approximation with one free particle is within 1\% of available simulation results. 
	The second term (two free particles) accounts for the remaining discrepancy \cite{Yam13}.
	This underlines the rapid convergence of the Stillinger series and assures us that the first term already contains the basic physics.
	Due to the short range of the HS potential, only the particles in the first shell around the vacancy
	contribute to the vacancy integral term ($\nmax = 1$ in \eq{eq:master_formula}).
	The corresponding results for $\nveq$ are shown in fig.~\ref{fig:HS_nvac} and show very good agreement with the simulation data above the
	liquid--solid coexistence density. The main contribution to $\dgv$ stems from the equation--of--state term $p/\rho \agt 10\, \kt$
	whereas the vacancy integral term contributes about $-1\, \kt$.

		\begin{figure}
			\centering
			\includegraphics[width=\picwidth]{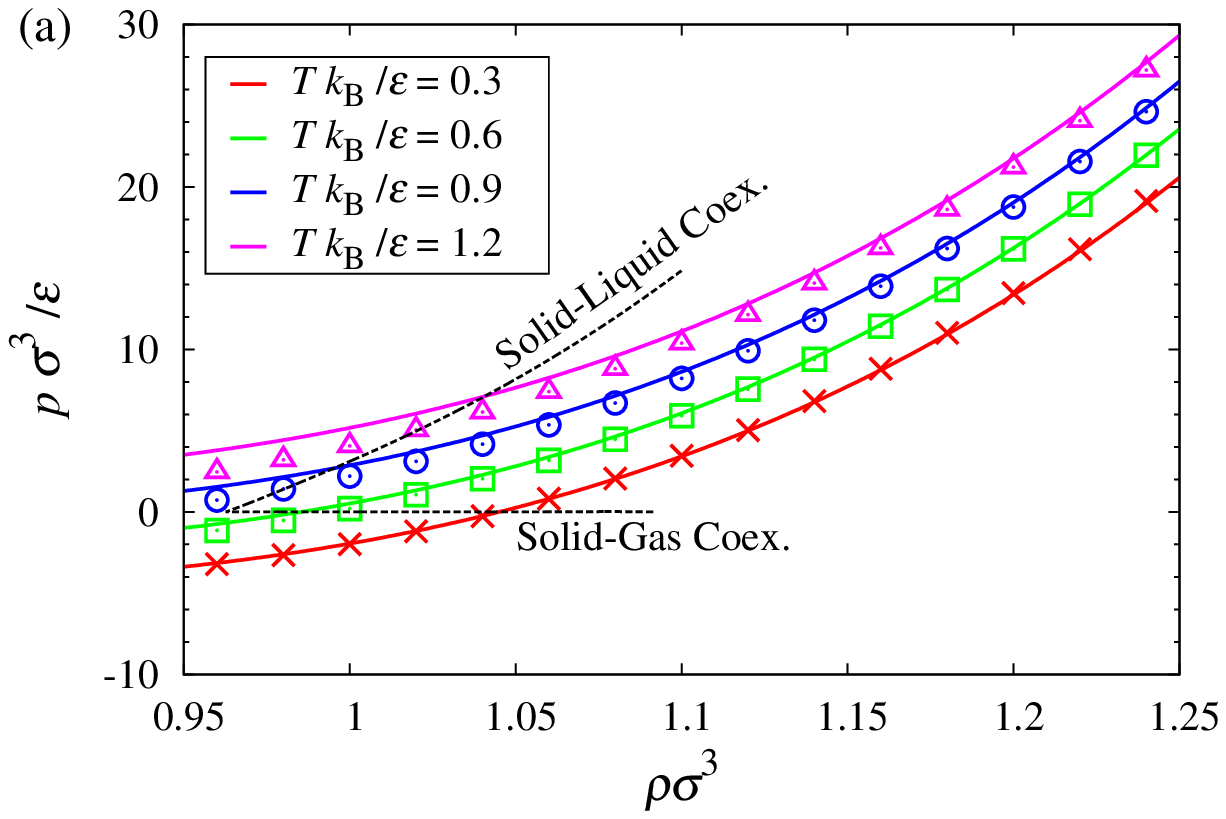}
			\includegraphics[width=\picwidth]{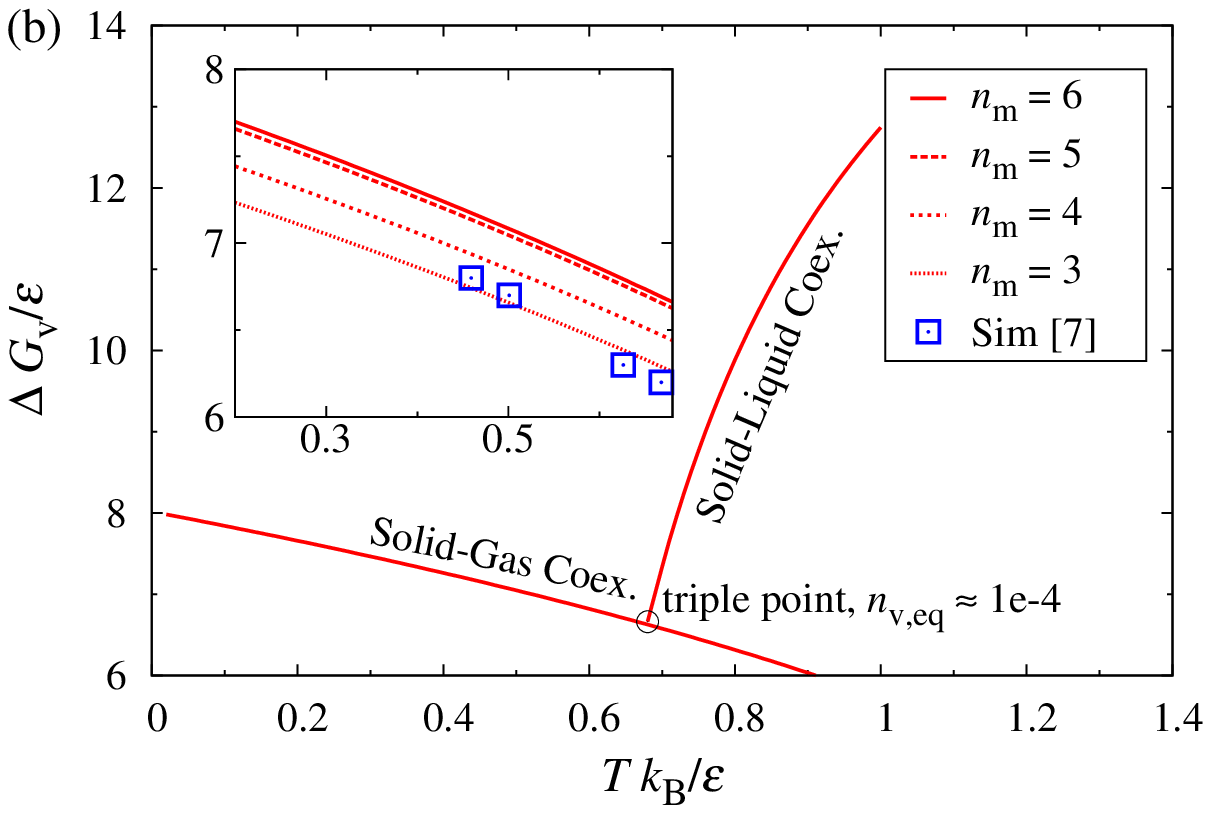}
			\includegraphics[width=\picwidth]{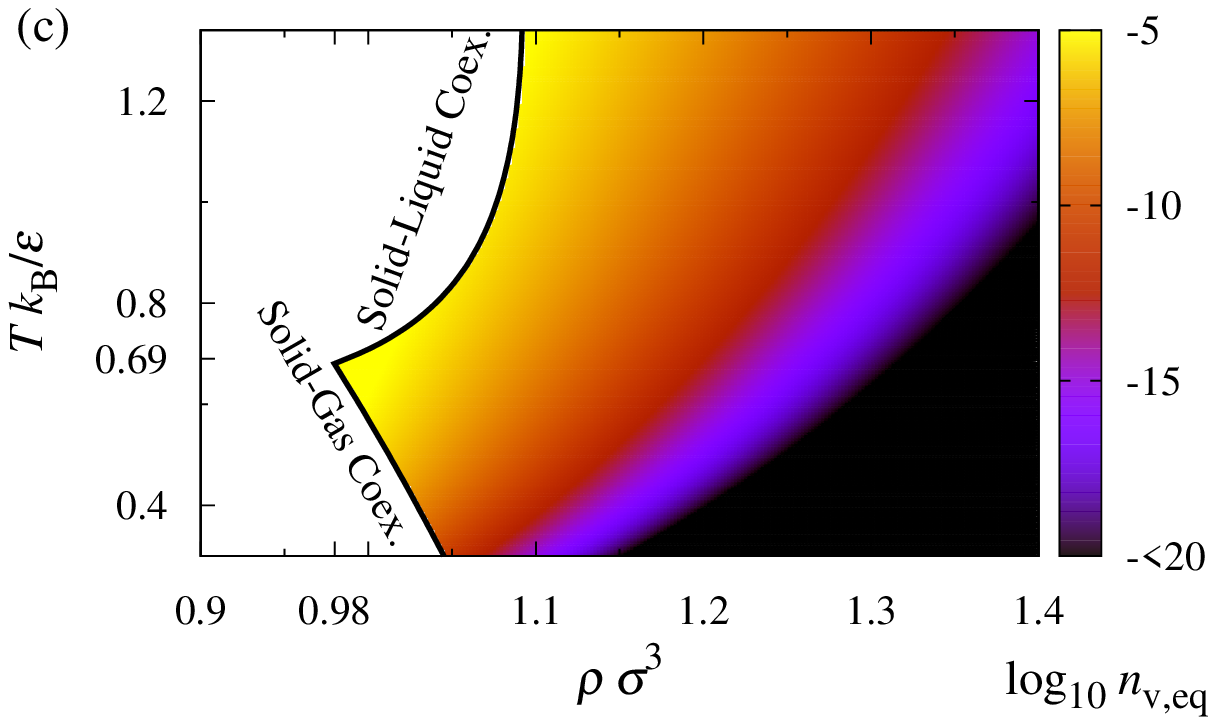}
			\caption{The Lennard--Jones system: (a) Equation of state, solid lines are the parametrizations of
			ref.~\cite{vdH00}, symbols are our data. (b) Gibbs energy of vacancy formation along the zero pressure line and the solid--liquid
			coexistence line, meeting in the triple point. The broken lines show the dependence of $\dgv$ on the potential
			range through the maximum shell index $\nmax$. Symbols are simulation data from ref.~\cite{Jac80}, error bars are unknown. 
			(c) Equilibrium vacancy concentration in the $\rho$--$T$ plane.
			We use the same long--range cutoff as in ref.~\cite{vdH00} ($r_c=6\,\sigma$). 
			  }\label{fig:LJ}
		\end{figure}

	In the case of LJ, we have used a truncated and shifted version of potential,
	$U_{LJ}^C(r) = U_{LJ}(r) - U_{LJ}(r_c)$ for $|r|< r_c$ and 0 otherwise, with
	$U_{LJ} = 4\epsilon\left\{\left(\frac{\sigma}{r}\right)^{12} - \left(\frac{\sigma}{r}\right)^6\right\}$ and $r_c$ is the cut-off range.
	The approximative crystal equation of state compares well with the parametrization of van der Hoef \cite{vdH00, vdH02}
	below and around the triple temperature (see fig.~\ref{fig:LJ}(a)). For higher temperatures deviations are visible, but at the same time also the liquid--solid
	coexistence density shifts upwards such that the stable crystal is still well described.
	In  fig.~\ref{fig:LJ}(b), we show $\dgv$ along the sublimation line ($p=0$, solid--gas coexistence) up to the triple point
	where it forks into the  liquid--solid coexistence
	line and the line of zero pressure.
	Comparison is only possible to the sublimation line simulation data of ref.~\cite{Jac80} where a vacancy has been placed in the middle of a rather small cubic simulation box
	of side length $3a$ where $a$ is the side length of the fcc cubic unit cell (i.e., the simulation box contains 107 particles). This means that the cutoff $\nmax$  
	on the maximal number of shells
	in \eq{eq:master_formula} has to be chosen such that the maximum shell radius $\approx 1.5\,a$  (i.e.,  vacancies in the periodic images do not interact). 
	This is the case for $\nmax=4$ with maximum shell radius $\sqrt{2}a$. Our data are consistent with the simulation data and furthermore show that
	the often--made assumption $\Delta S_{\rm v}$ being $T$--independent holds only approximately.
	We also see that near the triple point $\dgv \sim 10\, \kt$ as it was for HS near coexistence. The origin, however, is completely different: the pressure term in
	\eq{eq:master_formula} does not contribute but the vacancy integral term gives rise to ``missing cohesive free energy''.  
	For $T=0$, $\dgv$ becomes the vacancy formation energy $E_{\rm v}$ which for a system interacting with pair potentials is near the modulus of the cohesive energy $|E_{\rm coh}|$
	(it would be exactly equal if the atoms remain fixed at their ideal lattice sites around the vacancy). It is interesting to note that the effect of the collective 
	particle displacements (to make $E_{\rm v}$ minimal) is contained in the vacancy integral term through a sum of single--particle displacements in the potential
	field of otherwise fixed atoms.  
	Finally, fig.~\ref{fig:LJ}(c) shows 
	$\nveq$ for stable crystals in the $T$--$\rho$--plane which is quickly calculated using \eq{eq:master_formula}. At the triple point $\nveq\approx 10^{-4}$ is maximal.

\begin{figure}
			\centering
			\includegraphics[height=\picwidth, angle=-90]{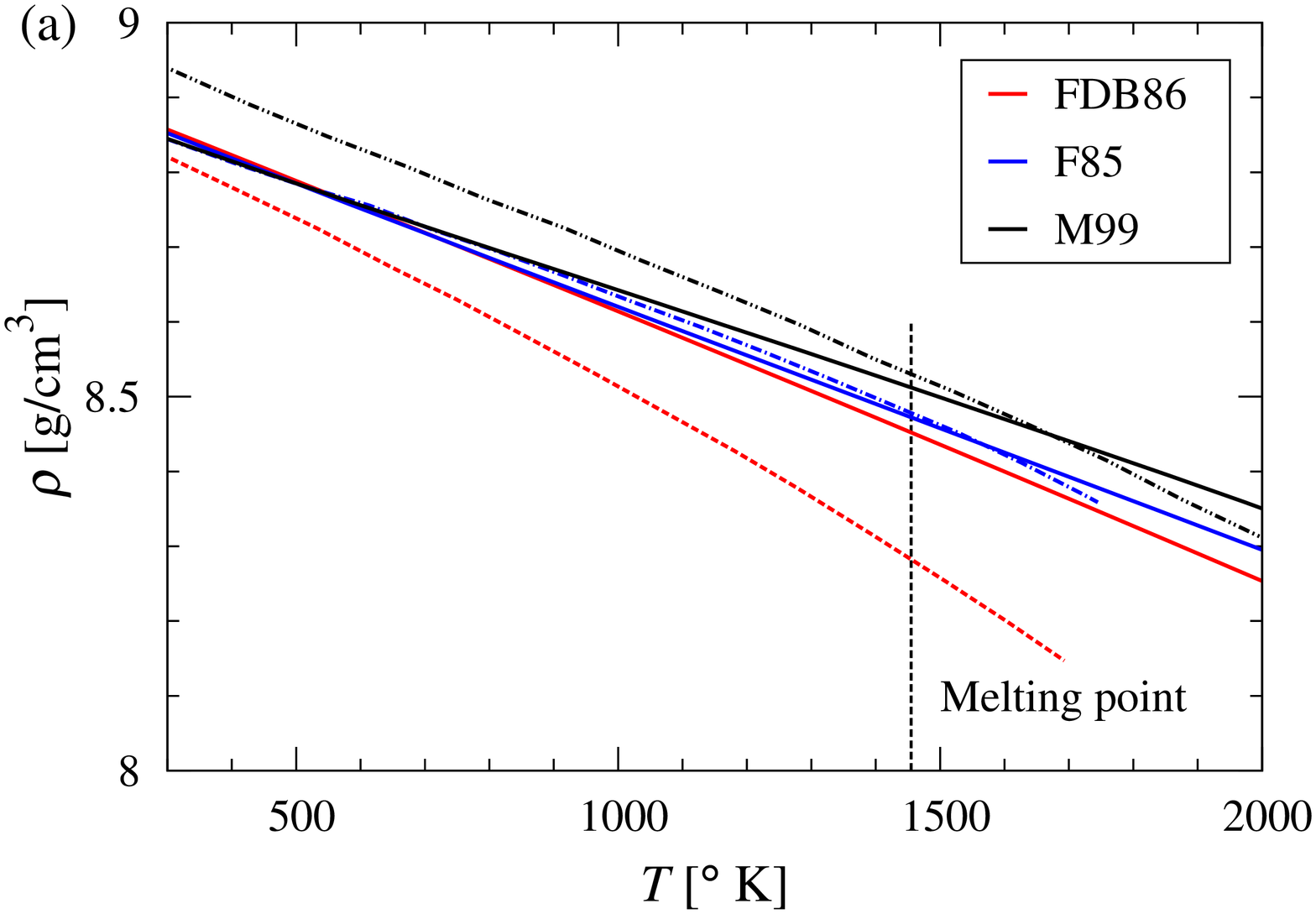}
			\includegraphics[height=\picwidth, angle=-90]{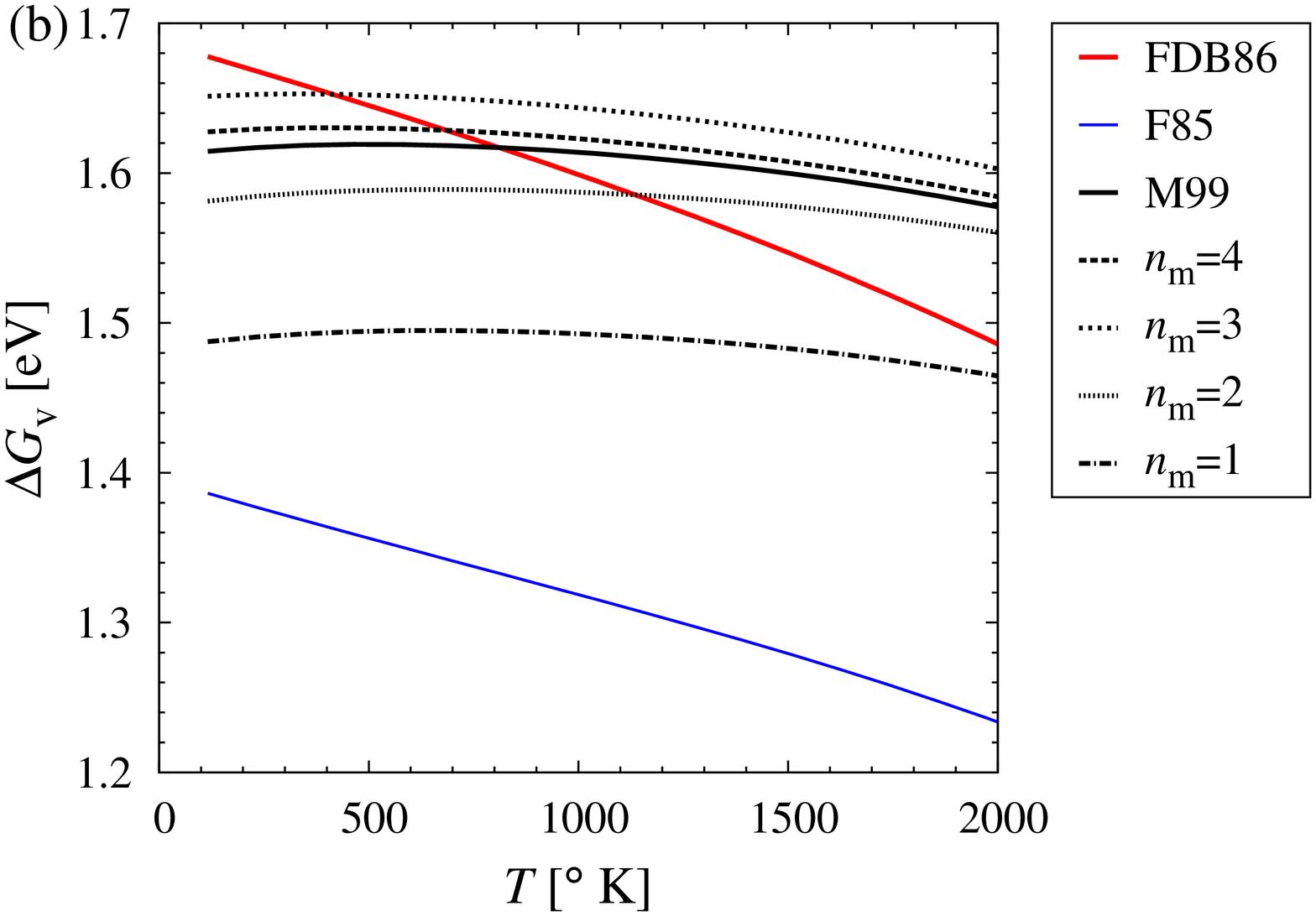}
			\caption{EAM results for Ni: (a) The variation of density with temperature along the zero pressure line.
			(b) Gibbs free energy of vacancy formation along the zero pressure line.
			For M99, the dependence on the maximum shell index $\nmax$ in the vacancy integral term  of \eq{eq:master_formula}
			is shown with broken lines. For F85 and FBD86 the behavior is similar to the LJ case (fig.~\ref{fig:LJ}(b)).
			Previously reported values at $T=0$ are $1.63$ eV for FBD86 \cite{FBD86},  $1.4$ eV for F85 \cite{Daw84}, and
			$1.6$ eV for M99 \cite{Mish99}. An experimental value is $(1.58-1.63) \pm 0.05$ eV \cite{Fel78}.
			}\label{fig:EAM_Ni}
		\end{figure}

	Next we turn to metals. In this case, the embedded-atom method (EAM) is a semi--empirical, classical many--atom potential for computing
	the total potential energy \cite{Foi93} and it can be used straightforwardly in \eq{eq:master_formula}. 
	With regard to vacancies, many--body effects are necessary to include in the description of metals since for zero temperature  $E_{\rm v} \approx 0.3 |E_{\rm coh}|$
	for a number of metals (and not $E_{\rm v} \approx |E_{\rm coh}|$ as for LJ) \cite{Foi93}.
	For the case of Ni, we examined three versions of the EAM potential: F85 from ref.~\cite{Foi85}, FBD86 from ref.~\cite{FBD86}, 
	and M99 from ref.~\cite{Mish99}.
	All of the potentials have been optimized with respect to a number of solid properties at $T=0$.
	{F85 and FBD86 contain a separation into two parts, a repulsive pair potential due to host nuclei which
	mimics the repulsion between charge--screened nuclei, and an embedding potential arising from the background electron density.
	For M99 however, such a physical interpretation is not intended, and a compound set of experimental and {\em ab initio} 
	simulation results are used for optimizing the potential parameters.}
	In fig.~\ref{fig:EAM_Ni}(a) we show
	the variation of the Ni density with temperature at zero pressure (sublimation line).
	Experimental data are not known but the agreement with simulation data is very good for F85 and M99
	\cite{Hor13}. 
	In fig.~\ref{fig:EAM_Ni}(b) we compare $\dgv$ for the three potentials.
	As a consistency check, at $T=0$ we recover the values for $\dgv=E_{\rm v}$ calculated previously in \cite{Daw84,FBD86,Mish99}
	which confirms that $E_{\rm v}$ is accurately calculated by summing over single--particle displacements.
	However, the temperature dependence of $\dgv$ is completely different for the two types of potentials  (F85/FBD86 vs. M99).
	$\dgv(T)$ is monotonously decreasing with increasing $T$ for F85/FBD86 with a slight increase of the slope for higher $T$.
	This is similar to $\dgv(T)$ for LJ (see fig.~\ref{fig:LJ}(b)) and also consistent with EAM potential molecular simulation
	data for copper (also with an fcc lattice structure) \cite{Foi94}. On the other hand, $\dgv(T)$ for M99 has a local maximum
	for $T \approx 500 K$. Furthermore, the variation of $\dgv$ with the maximum shell index $\nmax$ at a fixed temperature is also 
	non--monotonic, in contrast to F85/FBD86 and the Lennard--Jones solid.
	These results  
	reflect an essential difference in the parametrization especially of the pair part of the EAM potential.
	In F85  and FBD86, this pair part is purely repulsive and short--ranged (up to the first [F85] and third shell [FBD86]).
	M99 assigns a good part of the attractive energy to the pair part and the pair part is significantly longer--ranged 
	({cut off at the fifth shell}) and contains oscillations. These oscillations account for both the non--monotonicity of $\dgv$ in temperature 
	and $\nmax$.
	Thus finite--temperature behavior in observables is susceptible
	to the parametrizations of EAM potentials and should be considered from the beginning. 

{\em Summary and conclusion.--}
	We have shown that a single--cell approximation (leading term truncation of the Stillinger expansion for the partition function of a crystal) 
	gives a good description of the equation of state for the hard sphere and Lennard--Jones model systems as well as for the exemplary case of Ni in the
	embedded atom method. In the same truncation, Gibbs energies of vacancy formation $\dgv$ and corresponding equilibrium vacancy concentrations 
	have been calculated which show good agreement with simulations in the hard sphere and Lennard--Jones case. The compact expression for $\dgv$
	(\eq{eq:master_formula}) allows for a transparent interpretation of the two sources of contributions to $\dgv$ (finite pressure and missing cohesive
	{\em free} energy of particles near a vacancy). For the more difficult case of metals we have shown that the temperature dependence of $\dgv$ is 
	significantly affected by details of the EAM potential parametrization. Therefore we propose that $\dgv(T)$ should be included in further EAM parametrizations.
	As a reference, $\dgv(T)$ should be calculated in quantum
	density functional theory (qDFT) using \eq{eq:master_formula} with qDFT values for the potentials $\phi(1...N)$ needed in the single--particle
	integrals $Z_s$ and $Z_{sv,i}$. Furthermore, the partition function, free energy and equilibrium vacancy concentration in this truncation could be {\em measured}
	directly in experiments with colloidal crystals if fixing of all but one particle on a lattice can be realized. This fixing appears to be possible
	employing laser tweezers \cite{Gri97,Roy13}. The trajectory of the remaining free particle can be recorded and binned using videomicroscopy, thus directly accessing   
	the integrands of the single--particle
	integrals $Z_s$ and $Z_{sv,i}$ \cite{Bon12}.  
 
{\bf Acknowledgment:}
	The authors thank the German Research Foundation (DFG) for support within the priority program SPP 1296 ``Heterogeneous nucleation" (Oe 285/1-3).


\end{document}